\def\keV{{\rm\thinspace keV}}
\def\km{{\rm\thinspace km}}

\def\Msun{\hbox{$\rm\thinspace M_{\odot}$}}

\def\s{{\rm\thinspace s}}
\def\yr{{\rm\thinspace yr}}

\def\kmps{\hbox{$\km\s^{-1}\,$}}

\def\Msunpyr{\hbox{$\Msun\yr^{-1}\,$}}

\documentclass[11pt,twoside]{article}
\usepackage{newpasp,epsf}
\def\edcomment#1{\iffalse\marginpar{\raggedright\sl#1\/}\else\relax\fi}
\reversemarginpar

\begin{document}
\title{The interaction of radio sources and cooling flows}
\author{A.C. Fabian}
\affil{Institute of Astronomy, Madingley Road, Cambridge CB3 0HA, UK}

\begin{abstract} 
The X-ray emission in many clusters of galaxies shows a central peak
in surface brightness coincident with a drop in temperature. These
characterize a cooling flow. There is often a radio source also at the
centre of such regions. Data from Chandra now enables us to map the
interaction between the radio source and the intracluster medium.
Preliminary work shows no sign of heating of the gas beyond the radio
lobes, which are often devoid of cooler gas and so appear as holes.
In the case of the Perseus cluster around 3C84, the coolest X-ray
emitting gas occurs immediately around the inner radio lobes.
\end{abstract}

\section{Introduction}

The X-ray surface brightness continues to rise towards the centre in
two thirds of all clusters of galaxies. The radiative cooling time of
the gas there continues to decrease, dropping below 10~Gyr at
70--150~kpc and 1~Gyr at 5--30~kpc. The smaller radius values apply to
low luminosity clusters such as that in Virgo, the larger ones to the
high luminosity ones such as that in Perseus. Within this inner region
the gas temperature also drops, from 3--10~keV down to 1--3~keV, as
determined by simple single-temperature spectral fits. The gross
appearance of these regions fits that of a cooling flow (see Fabian
1994 for a review), in which radiative cooling of the central dense
parts of a hydrostatic atmosphere causes it to subsonically slump
inward under the influence of gravity and the weight of overlying gas.
About 10 per cent of clusters, such as the Coma cluster, do not show
the central peak in X-ray emission and have no cooling flow. Another
10--20 per cent of clusters, such as A754, have complex structure,
probably due to a cluster-cluster merger.

The inferred rate at which gas is cooling out in cooling flows ranges
from $<10\Msunpyr$ in poor clusters to $>1000\Msunpyr$ in massive rich
clusters. These values are uncertain and depend on an assumed age for
the flow if determined only from x-ray imaging. Only when X-ray
spectra show the cooling components can the rate be clearly defined.
In that case an estimate for the age of the flow is obtained as well.
Work using ROSAT and ASCA data (Allen \& Fabian 1997; Allen et al
1999) indicates that the ages may be about 5~Gyr. Many clusters appear
to have rates of a few $100\Msunpyr$, which suggests total cooled gas
masses of $10^{11} - 10^{12}\Msun$. Blue light from massive stars and
emission line nebulosities are common in such regions (see Crawford et
al 1999 and references therein) but normal star formation accounts for
only about 10 per cent of the cooling rate. 

What happens to the rest is not clear. It has been suggested that
heating balances cooling so that little or no gas actually cools from
the X-ray emitting hot gas component (Rosner \& Tucker 1989; Binney \&
Tabor 1995; Soker et al 2000). This seems plausible when it is
noted that most cooling flows have a radio source in the host central
cluster galaxy. If somehow the accretion power from a small amount of
gas accreted onto a central black hole can be distributed into the
cooler and cooling gas then perhaps some equilibrium may be obtained.
Thermal conduction too must be worried about since we are discussing
cooler gas components embedded in hotter ones.

The problem here is that the X-ray data show little sign that the
radio sources are actually heating the bulk of the gas. Of course
there is something going on in the radio lobes, but these occupy only
a small fraction of the flow. What is needed to stem a cooling flow is
a lot of energy into the cooler gas. It has to hit the gas with the
shortest cooling times. The heating needs to be complete in the sense
that it shouldn't just cause the gas to hang up at some temperature
below the outer cluster temperature or it would easily be detected. It
must also allow gas to cool (or give the appearance of cooling) down
to 1--3~keV. This is not trivial.

Here I review the recent results from Chandra which resolve the radio
-- X-ray interaction regions in some luminous clusters. The simple
result is that the coolest gas is seen to be closest to the radio
lobes. Not what a heating model would be expected to show if the lobes
are responsible for the heating.

\section{Chandra results}

\subsection{Hydra A and 3C295}

Early data from Chandra showed clear evidence for `holes' in the X-ray
in the vicinity of the radio lobes in the Hydra A cluster (McNamara et
al 2000). The central radio source here is fairly powerful and the
lobes occupy one of the larger fractions of the cooling region. Later
work suggests that a small cooling flow may operate only near the
centre of the region (David et al 2000) at a rate compatible with
the star formation seen.

The radio lobes in the powerful FRII source 3C295 are seen in X-ray
emission, presumable as a consequence of inverse-Compton scattering
(Harris et al 2000). The surrounding emission does however appear to
be reasonably undisturbed and undergoing a cooling flow (Allen et al
2000). 

\subsection{The Perseus cluster and 3C84}

The core of the Perseus cluster around the central galaxy NGC1275 and
its radio source 3C84 has long been known to have X-ray `holes' at the
positions of the radio lobes (Bohringer et al 1993). ROSAT data showed
good agreement between the holes and inner lobes. Chandra imaging now
shows this very clearly (Fabian et al 2000a; Fig. 1). 

\begin{figure}
\plotone{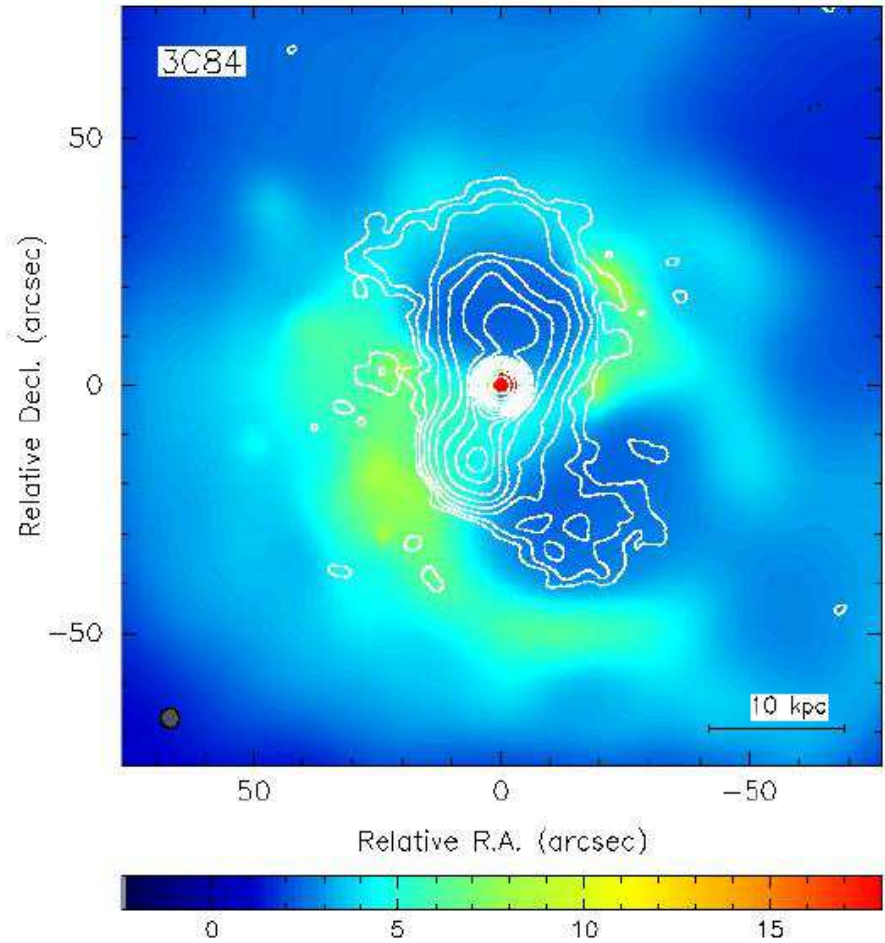}
\caption{Radio image (1.4~GHz restored with a 5 arcsec beam, produced
by G. Taylor; see Fabian et al 2000a) overlaid on adaptively smoothed
0.5--7~keV X-ray map.}
\end{figure}

The exact 3D geometry is not obvious (Fig. 2). Presumably the S jet is
pointing toward us at some angle. There are also several outer holes
seen (Fabian et al 2000a) which may represent older, buoyant lobes
(Churazov et al 2000a). The energetic electrons may now have cooled in
these regions but the nonthermal pressure from the protons,
magnetic field and cooled electrons may keep them inflated and less
dense than the surroundings. A spur of radio emission toward the W
outer hole is seen in the 74~MHz map of Blundell et al (2000) which
supports this idea.

\begin{figure}
\plottwo{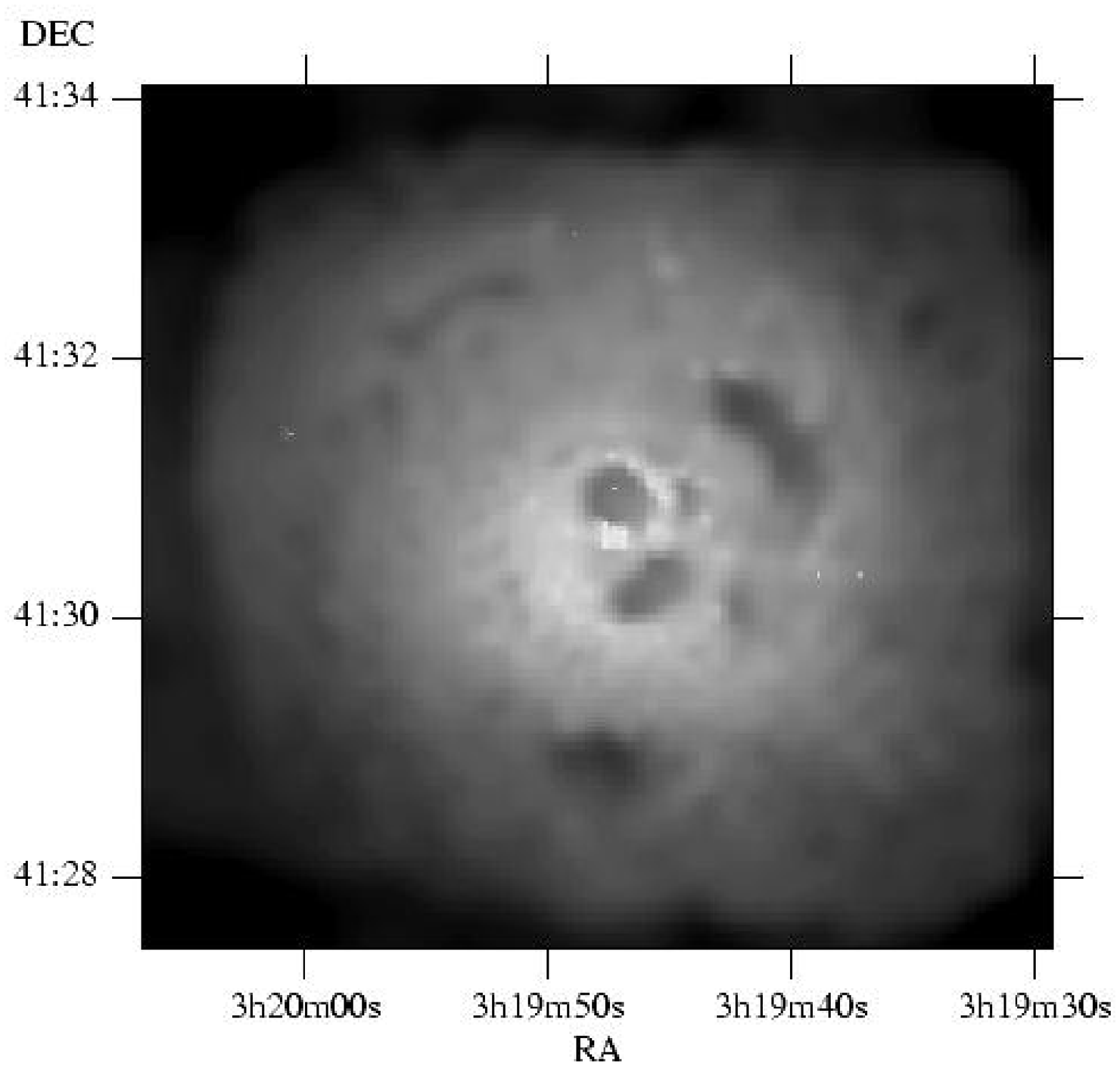}{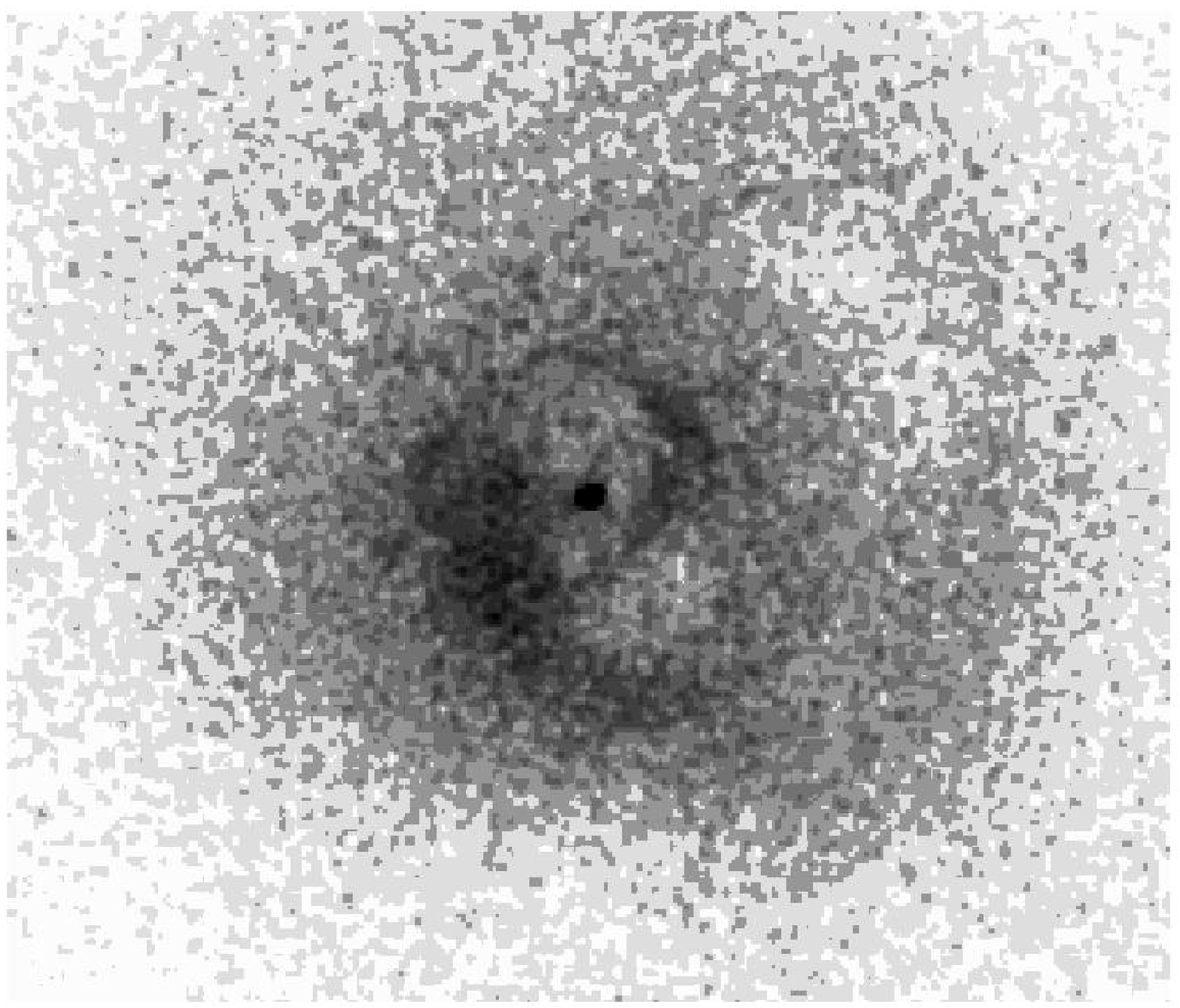}
\caption{(Left) Adaptively-smoothed 0.5--7~keV ACIS-S X-ray image; (Right)
gaussian-smoothed ACIS-I X-ray image of the centre of the Perseus
cluster. The ACIS-S image has cosmetic blurring due to a nodal
boundary running from just below the nucleus to the W. These effects
are absent in the I image. Note that the I image shows better the
complex structure of the inner part (compare with Fig.~1). The precise
nature of this structure is not clear. The holes may be roughly in the
plane of the Sky or, more likely, arranged in a perpendicular sense
with the apparent `e' shape being part of a helix oriented along our
line of sight. }
\end{figure}

\begin{figure}
\plottwo{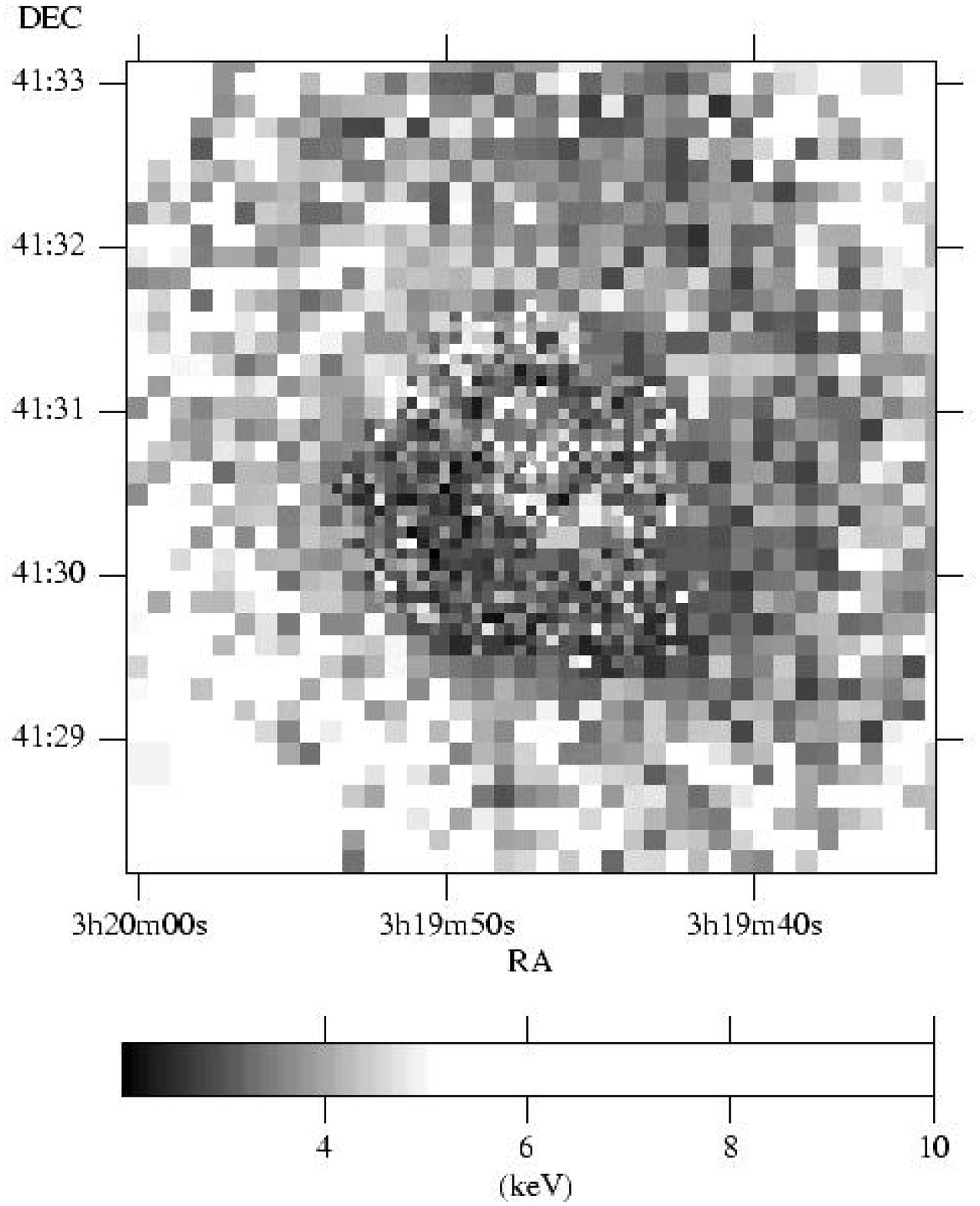}{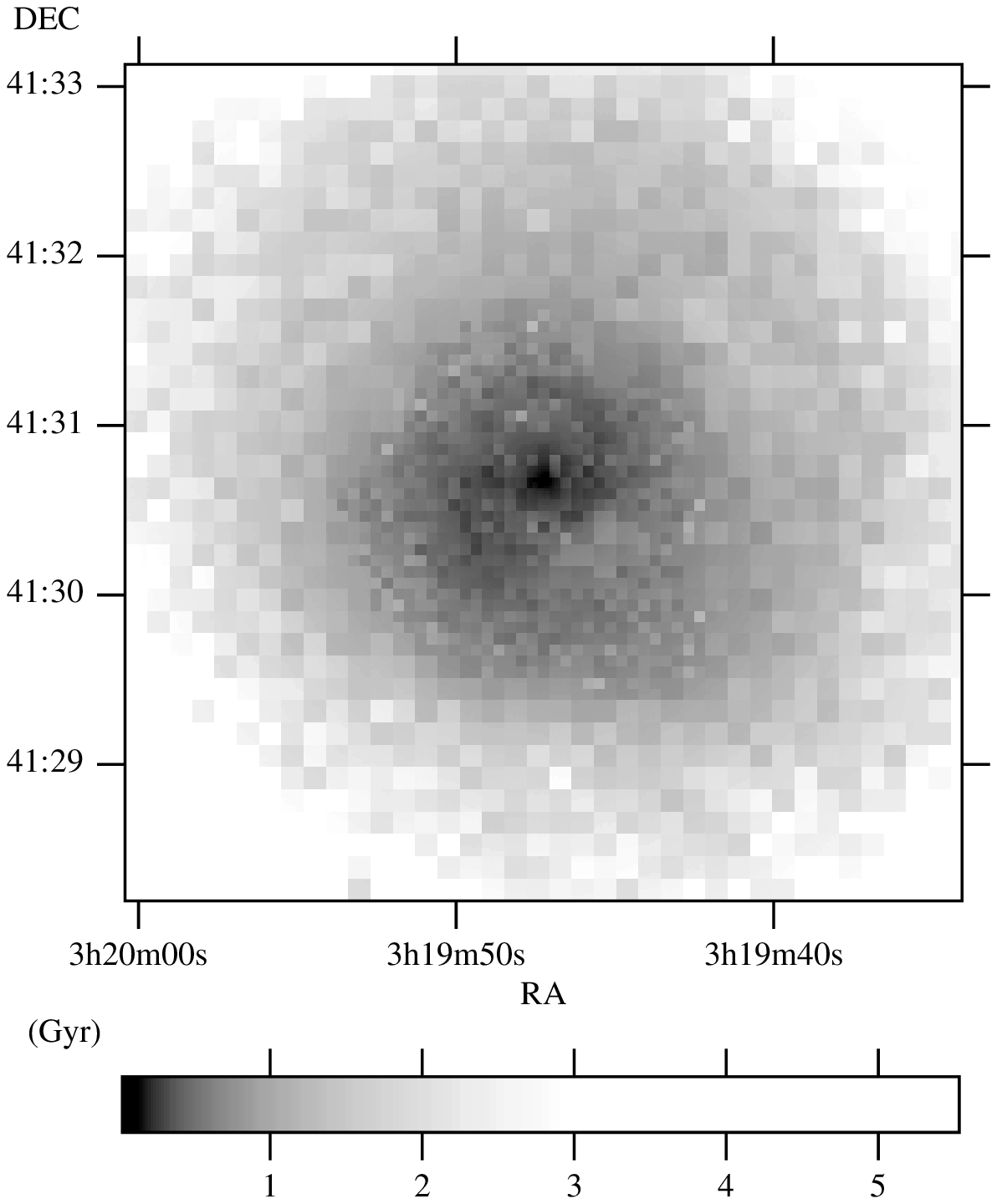}
\caption{(Left) Temperature and (Right) radiative cooling time maps
produced from the X-ray colour ratios (Fabian et al 2000a). Note that
the coolest gas ($T\sim 2.5\keV$) with the shortest cooling time
($\sim 0.3$~Gyr) lies in the rim around the N lobe and in the E bright
blob. Single-phase gas has been assumed for the analysis. }
\end{figure}

The Chandra data (Fabian et al 2000a) covers the 0.3--7~keV range and
can be divided into colours from which absorption and temperature maps
can be constructed. Significant excess absorption is seen in the
region of the `high velocity system', an irregular galaxy falling into
the cluster core close to our line of sight. The temperature map (Fig.
3) shows that the gas is cooler toward the centre with the coolest gas
along the rims of the inner radio lobes and in the bright E blob. The
surface brightness can be used to obtain the gas density and thus
pressure and radiative cooling time. The cooling times are also
shortest close to the inner radio lobes.

We see no sign that the lobes have heated the gas beyond the lobes.
Something has happened within the lobes. They have either been cleared
of gas, or perhaps just cleared, of cooler gas. The intracluster medium
there may be multiphase with cooler, cooling, blobs embedded within a
hotter phase. The lobes may have only pushed aside the cooler blobs. A
simple rearrangement of the cooler gas is in agreement with the
observed surface brightness excess around the rims of the lobes. There
is certainly no sign of the strong shocks suggested by Heinz et al
(1998).

\subsection{A1795 and A2199}

\begin{figure}
\plottwo{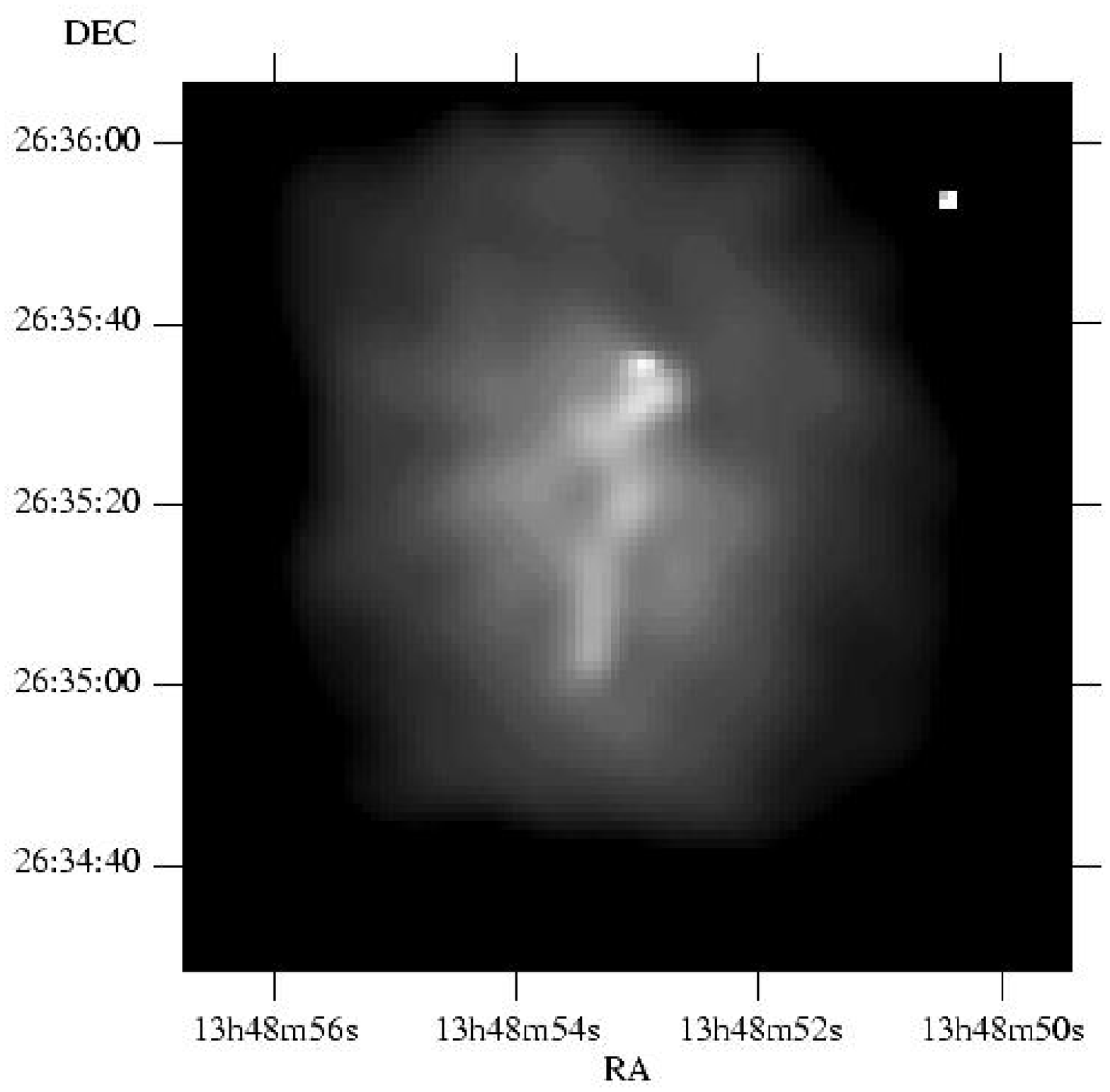}{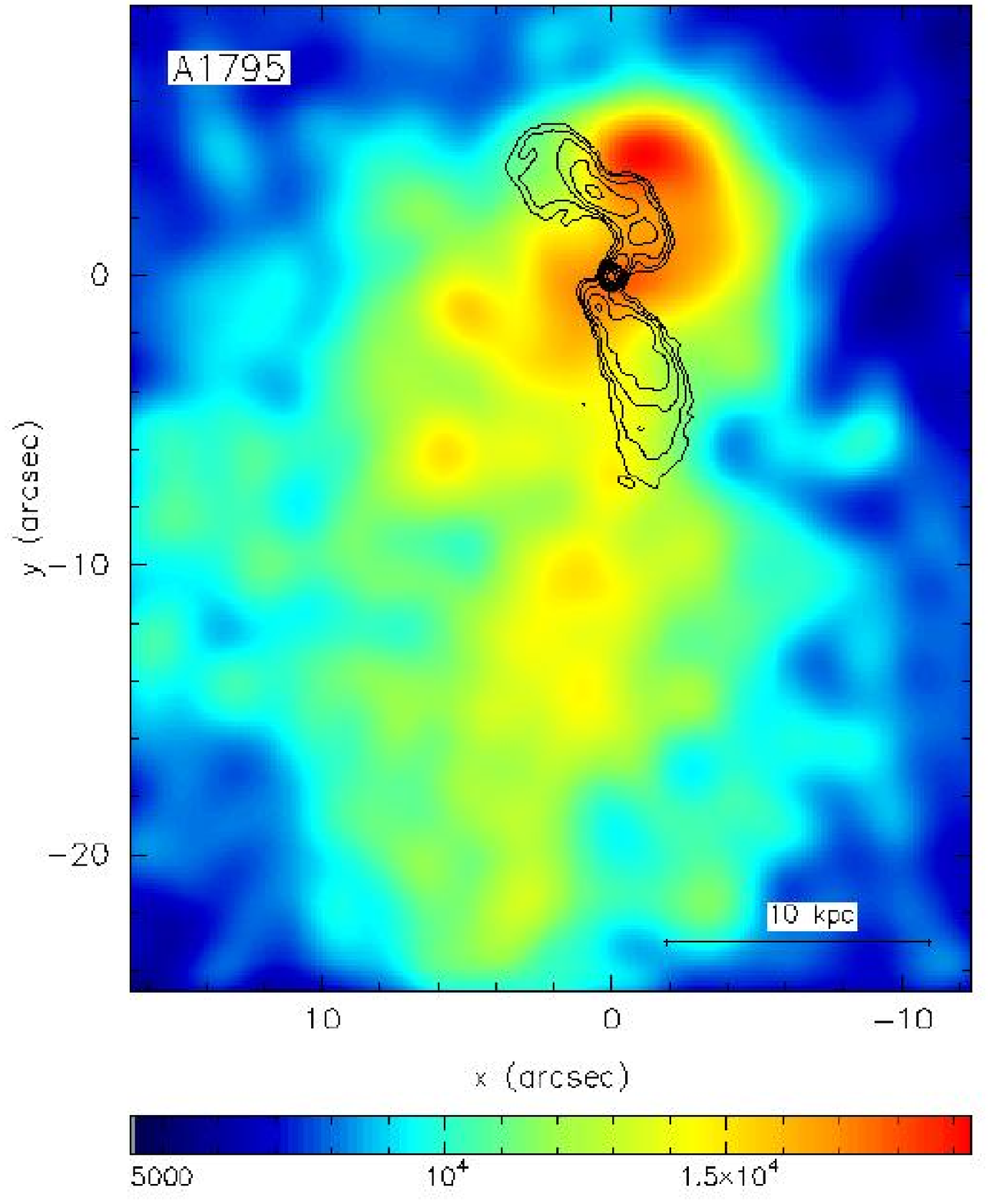}
\caption{(Left) Adaptively-smoothed X-ray image of the centre of A1795
(Fabian et al 2000). (Right) Overlay of the 3.6~cm radio emission (Ge
\& Owen 1993) on the X-ray image.}
\end{figure}

Other clusters with central radio sources which we have observed are
A1795 (Fabian et al 2000) and A2199 (Johnstone et al 2001, in
preparation). A1795 shows a filament of soft X-ray emission (Fig. 4)
which coincides with an optical H$\alpha$ filament discovered by Cowie
et al in 1983. This could be a cooling wake due to motion of the
central cluster galaxy. The velocity map of the H$\alpha$ emission (Hu
et al 1985) shows that the wake has the same velocity as the bulk of
the cluster (Oegerele \& Hill 1994), there is a velocity shift of
$+150\kmps$ at the cD galaxy. The small radio source associated with
the cD (Ge \& Owen 1993) shows some interaction with the brighter
local X-ray emission. A possible X-ray `shadow', a fainter X-ray
region, parallel to but further out from the radio lobes, is also seen. 

A2199 has large `holes' conicident with its more diffuse outer radio
structures (Fig. 5; radio data from Giovannini et al 1998, see also
Owen \& Eilek 1998) but there is little agreement between the inner
radio and X-ray structures. Analysis of the Chandra data on A2199 is
in progress (Johnstone et al 2001).

\begin{figure}
\plottwo{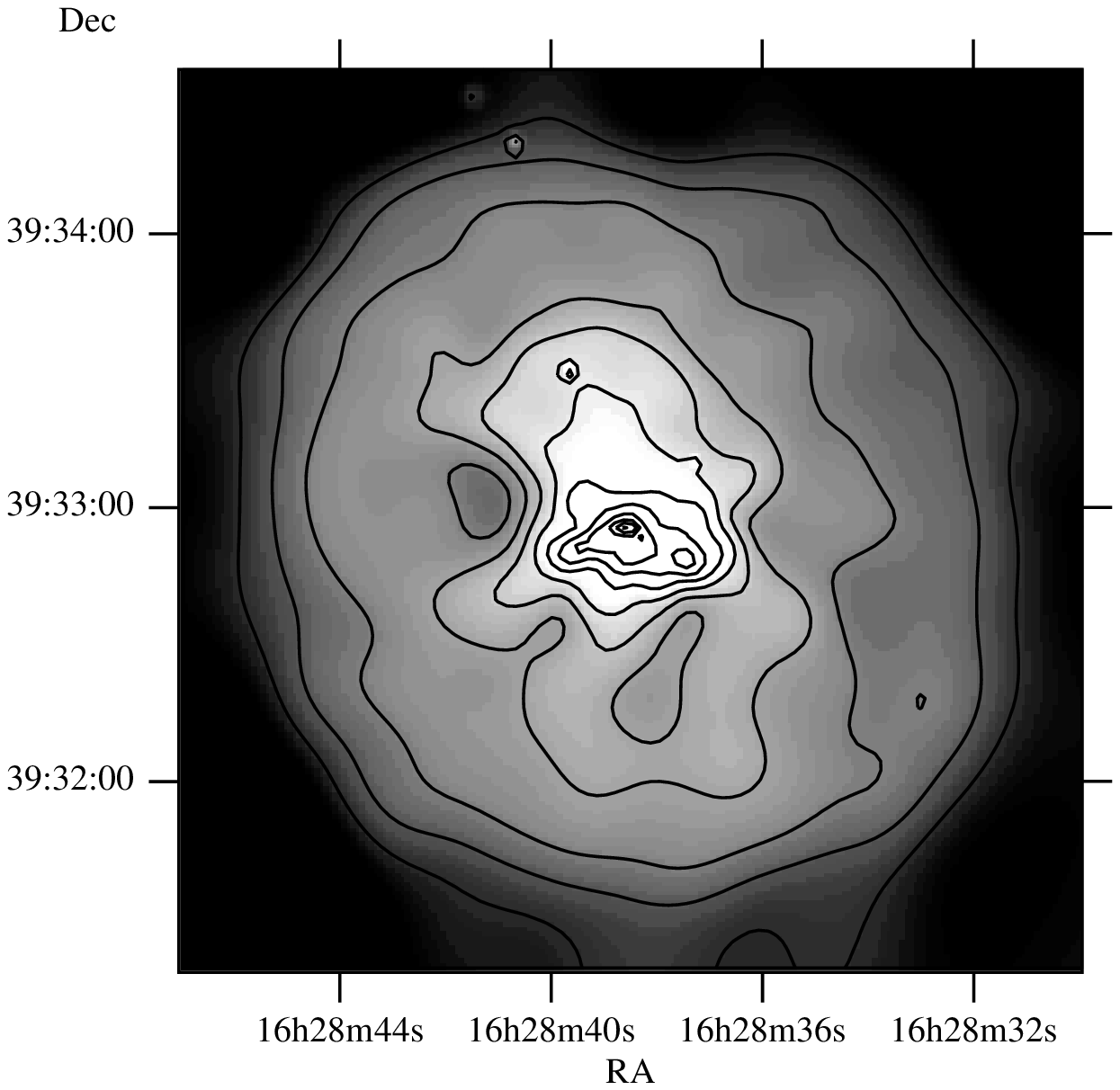}{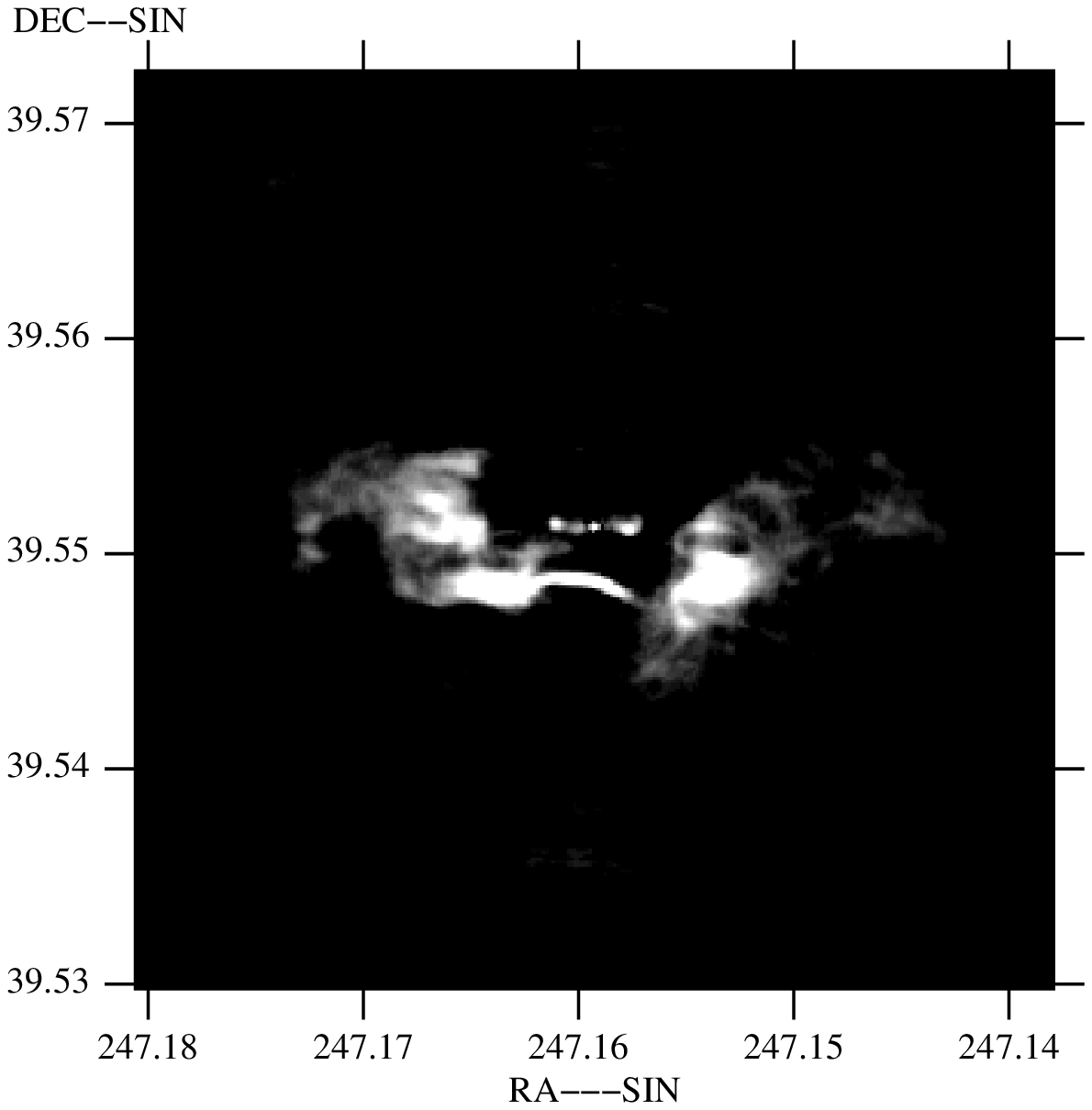}
\caption{(Left) Adaptively-smoothed 0.5--3~keV X-ray image of the core
of A2199 (the contours are logarithmic); (Right) 1.7~GHz radio image
(Giovannini et al 1998). }
\end{figure}

\subsection{The Virgo cluster and M87}

\begin{figure}
\plotone{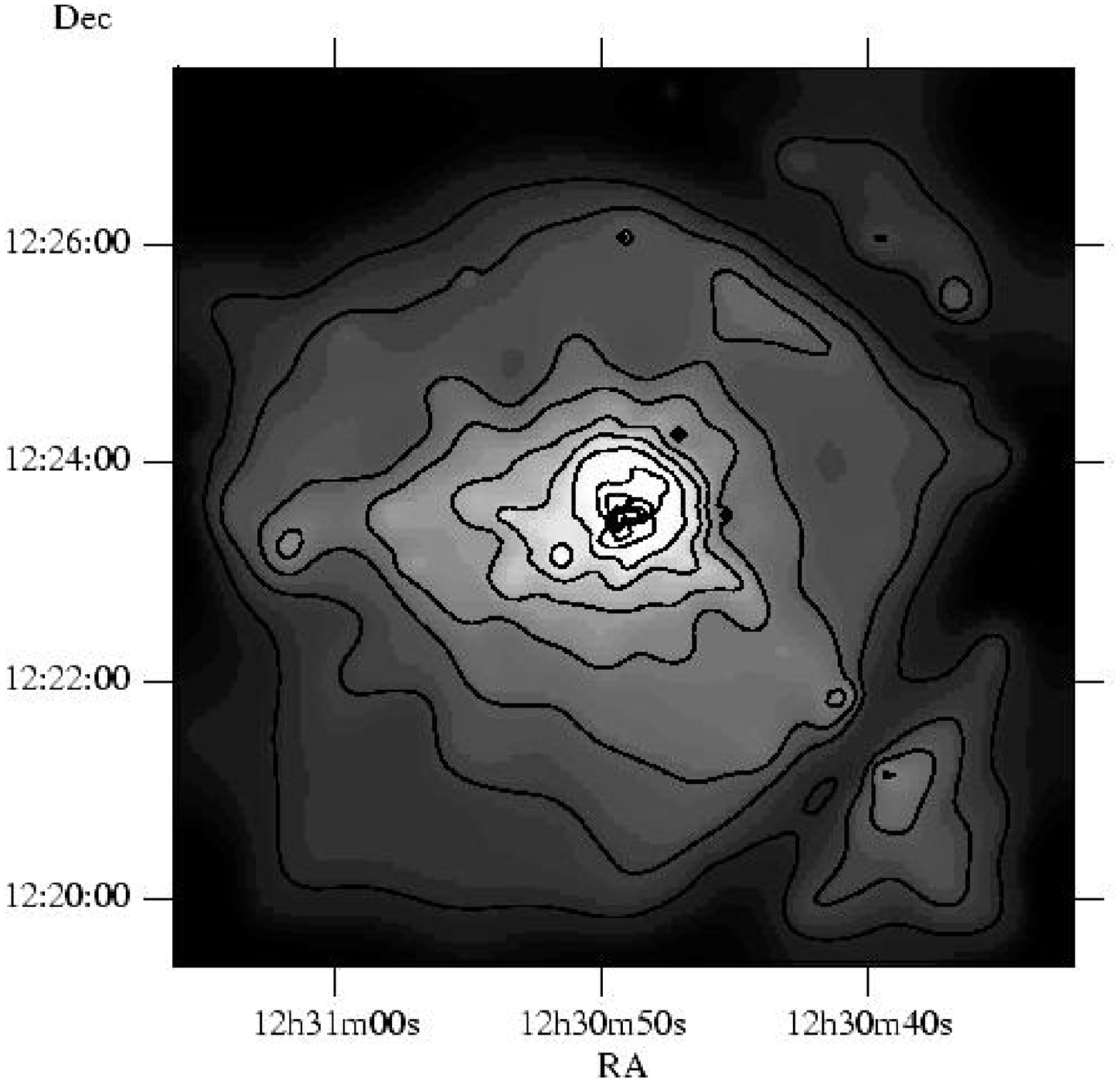}
\caption{Chandra ACIS-I image (0.5--2.5~keV) of the centre of the
Virgo cluster around M87. The dark diagonal structures are due to chip
gaps. }
\end{figure}

Chandra images of the inner parts of the Virgo cluster around M87 have
been made in the ACIS-I detector by Fabian et al (2001 in preparation)
and in the ACIS-S detector by Wilson et al (2001). The excess emission
along the line of the radio structures, seen in ROSAT data (Bohringer
et al 1995), is clearly detected (Fig.~6). Only the X-ray structure to
the E shows any good correspondence with the radio images (Owen et al
2000). The excess emission there is thermal and cooler than the
surrounding gas (Bohringer et al 1995) and possibly due to uplift of
cooler gas from nearer the centre by a buoyant radio plume (Bohringer
et al 1995; Churazov et al 2000b).

\section{Summary}

The Chandra data so far show little evidence for any widespread steady
heating by the radio source. Perhaps there are sporadic upheavals due
to large outbursts by the central engine, or a pervasive leakage of
energetic particles. Or perhaps cooling dominates.

Conduction may be highly suppressed in cooling flows. Chandran et al
(1999) have discussed theoretical possiblilities including the
mirroring of particles. The Chandra observations of `cold fronts' in
some clusters (Markevich et al 2000; Vikhlinin et al 2000) where 5 keV
gas lies a few kpc away from 10~keV gas suggests observationally that
conduction may in some intracluster regions be highly suppressed
(Ettori \& Fabian 2000).

\section{Post-conference appendix}

In between the conference and the writing of this paper,
Reflection-Grating Spectrometer (RGS) data from XMM have appeared.
These have a strong bearing on the appearance of cooling flows so a
brief outline of the issues is now given to steer the reader to the
relevant literature.

The RGS data of the strong, hot, cooling flow cluster A1835 (Peterson
et al 2000) show none of the strong emission lines expected from gas
cooling below 3~keV. It seems like the gas cools from about 9~keV down
to about 3~keV then vanishes. Similarly there are no lines from gas
below about 2~keV from A1795 (Tamura et al 2000) or A1101 (Kaastra et
al 2000). The temperature profiles of the gas drop inward in the manner
expected from a cooling flow but the spectral signature of the final
temperature plunge is missing.

There are many possible explanations, discussed by Peterson et al
(2000) and Fabian et al (2000c). Perhaps there is heating. But why then
does the gas apparently cool most of the way and is then all heated
back up to its original temperature? (It cannot be heated to some
other intermediate temperature or it would be detected as accumulating
there.) Perhaps the ages of the flows are all only about one Gyr.
Perhaps resonance scattering and some absorption operate on the inner,
cooler gas making it undetectable. Perhaps the metallicity of the gas
is bimodal. A recent paper on M87 by Bohringer et al (2000) shows that
the situation is complex, and metallicity, resonance scattering or
continuum dilution may be factors. They find that the metallicity
gradient in the Virgo cluster in which the iron (and other element)
abundance rises inwards reverses in the innermost arcmin resulting in
the metallicity going to zero! Not only are no lines seen from cooling
gas, but there are no lines from anything else. Clearly further
observations, and thought, are required.

\end{document}